\documentclass[twocolumn]{aastex62}
\usepackage{textcomp}
\usepackage{amsmath}     
\usepackage{wasysym}           
\usepackage{graphicx}
\usepackage{amssymb}
\usepackage{epstopdf}
\usepackage{mathrsfs}
\usepackage{anyfontsize}
\usepackage{natbib}
\usepackage{placeins}
\usepackage{color}
\usepackage{lipsum}
\usepackage{diagbox}
\usepackage{gensymb}
\usepackage{booktabs}
\usepackage{appendix}
\usepackage{graphicx}
\usepackage[left]{lineno} 

\usepackage{CJKutf8}

\shorttitle{the environment in LSB and HSB}
\shortauthors{Shen et al.}
\begin{document}
\begin{CJK}{UTF8}{gbsn}	
    
    \title{Beyond Mass and Multiscale Environments: What Shapes Low Surface Brightness Galaxies? Evidence from MaNGA}

    \correspondingauthor{Hassen M. Yesuf, Chong Ge}
    \email{yesufh@shao.ac.cn, chongge@xmu.edu.cn}
    
    \author[0009-0003-2072-1200]{Mengting Shen}
    \affiliation{Department of Astronomy, Xiamen University, Xiamen, Fujian 361005, China}
    
    \author[0000-0002-4176-9145]{Hassen M. Yesuf}
    \affiliation{Key Laboratory for Research in Galaxies and Cosmology, Shanghai Astronomical Observatory, Chinese Academy of Sciences, 80 Nandan Road, Shanghai 200030, China}
    
    \author[0000-0003-2478-9723]{Lei Hao}
    \affiliation{Key Laboratory for Research in Galaxies and Cosmology, Shanghai Astronomical Observatory, Chinese Academy of Sciences, 80 Nandan Road, Shanghai 200030, China}

    \author[0000-0003-0628-5118]{Chong Ge}
    \affiliation{Department of Astronomy, Xiamen University, Xiamen, Fujian 361005, China}
    
    \author[0000-0002-4499-1956]{Jun Yin}
    \affiliation{Key Laboratory for Research in Galaxies and Cosmology, Shanghai Astronomical Observatory, Chinese Academy of Sciences, 80 Nandan Road, Shanghai 200030, China}
    
    \author[0000-0003-4874-0369]{Junfeng Wang}
    \affiliation{Department of Astronomy, Xiamen University, Xiamen, Fujian 361005, China}
    
    \author[0000-0002-3073-5871]{Shiyin Shen}
    \affiliation{Key Laboratory for Research in Galaxies and Cosmology, Shanghai Astronomical Observatory, Chinese Academy of Sciences, 80 Nandan Road, Shanghai 200030, China}
    
    \begin{abstract}

    The origin of low surface brightness (LSB) galaxies remains a key open question in galaxy formation, reflecting the balance internal mechanisms and environmental influence. Using MaNGA integral-field spectroscopy, we investigate whether LSB and high surface brightness (HSB) galaxies of comparable stellar mass ($9 < \log M_\ast < 10$) occupy distinct environments or differ primarily through internal evolution. Our late-type sample comprises 113 central and 29 satellite LSB galaxies, and 374 central and 142 satellite HSB galaxies. We characterize environments on scales from $\sim$100 kpc to 10 Mpc, analyzing radial profiles of stellar mass surface density ($\Sigma_\ast$), star formation activity, and gas-phase metallicity. Central LSB and HSB galaxies inhabit similarly low-density large-scale ($>$200 kpc) environments, but LSB galaxies are more isolated on small scales ($\sim$100 kpc). Even after matching in stellar mass and environment, LSB galaxies show systematically lower $\Sigma_\ast$, $\Sigma_{\rm SFR}$, and metallicities, often hosting diffuse, weakly star-forming bulges embedded in extended disks. These results indicate that LSB structure and star formation are not primarily governed by large-scale environment or halo mass. While secondary halo properties such as spin, concentration, or gas accretion history are often invoked, their environmental dependence appears weak. Instead, LSB–HSB differences for centrals likely reflect divergent assembly or interaction histories and internal processes---such as angular momentum–driven disk evolution or inefficient gas conversion---largely decoupled from large-scale environment. Nonetheless, environment still influences the observed star formation and chemical differences between central and satellite LSB galaxies.

    \end{abstract}
    
    
    \section{Introduction}

    LSB galaxies --defined by disk central surface brightness $\mu_{0,d} \geq 22.5$ mag arcsec$^{-2}$, roughly one magnitude fainter than Freeman’s classical threshold \citep{Freeman_1970}-- represent a numerically significant but observationally elusive population. Although they contribute only a few percent to the local luminosity and stellar mass density \citep{Martin_2019}, LSB galaxies may account for $\sim$20\% of cosmic dynamical mass \citep{Minchin_2004} and 30–60\% of local galaxy number density \citep{O'Neil_2000}. Their defining characteristics---low stellar mass surface density, blue colors, suppressed star formation, and metal-poor gas---suggest slow evolution despite being exceptionally rich in neutral hydrogen (H{\sc i}), with H{\sc i} masses comparable to HSB galaxies \citep{Mo_1994, He_2020}. This arises from their extended dark matter halos, which stabilize disks and suppress conversion of atomic gas into stars \citep{de_Blok_1997, Perez_Montano_2022}, while low-density stellar disks further limit star formation efficiency. Consequently, LSB galaxies retain high gas fractions and remain dark matter dominated at all radii \citep{de_Blok_McGaugh_1996, Chequers_2016}.
    
    Observations and simulations consistently show that LSB galaxies have large H{\sc i} reservoirs, extended star formation histories, low S\'ersic indices, and slowly rising rotation curves \citep{Di_Cintio_2019, Perez_Montano_2024}. Crucially, their formation is primarily governed by the intrinsically high dark matter halo spin, which establishes an extended, low-density disk as the baseline structure \citep{Jimenez_2003, Kim_2013}. Event-driven processes, such as mergers or gas accretion, can subsequently redistribute angular momentum, modulating disk structure and surface brightness. Coplanar, co-rotating mergers or aligned gas accretion tend to preserve or enhance diffuse morphologies, whereas perpendicular mergers or misaligned accretion remove angular momentum, leading to more compact, higher-surface-brightness systems by $z=0$ \citep{Di_Cintio_2019, Kulier_2020}. Although halo spin and angular momentum alignment are dominant, other factors --such as halo concentration, feedback-driven outflows, and the timing of mergers-- seem to play only secondary roles \citep{Di_Cintio_2019, Perez_Montano_2024, Stoppacher_2025}. In this framework, the environment may influence the frequency of mergers or gas content, but has a limited effect on the intrinsic processes shaping LSB structure. These results support the ``halo spin'' model, where high-angular-momentum halos regulate disk scale lengths and produce LSB galaxies \citep{Dalcanton_1997, Boissier_Monnier_2003}, explaining their suppressed bar formation and steady evolution \citep{Cervantes-Sodi_2013}.
    
    An alternative hypothesis of "environmental control" proposes that LSB galaxies preferentially reside in low-density, isolated regions \citep{Mo_1994, Rosenbaum_2004}. Early studies supported this view, suggesting that such isolation limits galactic interactions and keeps gas densities low and suppresses star formation \citep{van_der_Hulst_1993, Pickering_1997}. However, more recent work shows that LSB galaxies are not preferentially isolated and their star formation is largely decoupled from large-scale density \citep{Galaz_2011, Shao_2015, Kulier_2020}. Simulations further indicate that local processes—such as mergers and gas accretion—dominate their evolution, while environmental isolation primarily acts as a protective factor that preserves their diffuse structure rather than driving their formation \citep{Di_Cintio_2019, Kulier_2020}. Thus, the environment is now regarded as a secondary influence that sustains, rather than creates, the distinct properties of LSB galaxies.
    
    Despite these insights, key gaps persist. Previous observational studies have not systematically separated central and satellite LSB galaxies in multiscale environments ($\sim$100 kpc–10 Mpc), nor compared their spatially resolved properties, such as $\Sigma_\ast$, metallicity and star formation distribution, to those of HSB galaxies matched in $M_\ast$ and multiscale environments.
    
    To address this gap, we present in this second paper of our series an analysis of LSB and HSB galaxies using integral-field spectroscopy from the Mapping Nearby Galaxies at Apache Point Observatory (MaNGA) DR17 \citep{Shen_2026}. Our sample consists of late-type galaxies within a stellar mass range of $9 < \log(M_\ast/M_\odot) < 10$, corresponding to the high-mass end of the distribution and thus representing only a subset of the LSB galaxy population. We characterize their environments on multiple scales and connect them to resolved internal properties to disentangle intrinsic drivers from environmental effects. Our results show that LSB and HSB galaxies inhabit broadly similar large-scale environments but differ on smaller scales, particularly among satellites. Even after matching in $M_\ast$ and several multiscale environmental indicators, LSB galaxies exhibit distinctly lower $\Sigma_\ast$, $\Sigma_{\rm SFR}$, and metallicities, and more extended star formation-consistent with intrinsic drivers. Although intrinsic processes dominate, the environment exerts a secondary influence on star formation and metallicity in satellite LSB galaxies.
    
    This work bridges theoretical models and observational constraints, providing a unified framework for how internal mechanisms and environment jointly shape galaxy diversity. We describe our sample, methods, and results below, and discuss their implications for LSB formation. We adopt a flat $\Lambda$CDM cosmology with $H_0 = 70$ km s$^{-1}$ Mpc$^{-1}$, $\Omega_{m} = 0.30$, and $\Omega_{\Lambda} = 0.70$ throughout.
    
    \section{DATA AND SAMPLE}
    \label{sec:DATA AND SAMPLE}
    
    The MaNGA survey, a core SDSS-IV project, has provided spatially resolved spectroscopy for $\sim$10,000 nearby galaxies ($0.01 < z < 0.15$; 10$^9$–10$^{11}$ M$_\odot$, \citeauthor{Blanton_2017} \citeyear{Blanton_2017}). Using 17 IFUs with a spatial resolution of 2$\arcsec$ ( $\sim$1.2 kpc) and covering 3,600–10,000 \AA, with a spectral resolution of $R \sim 2000$, it enables detailed analyses of emission lines and 3D spectral data \citep{Wake_2017,Law_2016}. In this study, we use MaNGA Product Release 11 (MPL-11), which contains 10,782 unique galaxies processed through both the Data Reduction Pipeline (DRP; \citeauthor{Law_2016} \citeyear{Law_2016}) and the Data Analysis Pipeline (DAP; \citeauthor{Westfall_2019} \citeyear{Westfall_2019}). These data provide spatially resolved measurements of star formation, metallicity, emission-line fluxes, and spectral indices, enabling detailed analysis of radial profiles and physical processes shaping galaxy evolution.
    
    Our parent sample is drawn from 1,118 face-on, late-type, non-AGN galaxies in SDSS–MaNGA DR17, for which bulge–disk decompositions are performed in the $g$ band using GALFIT software \citep{Peng_2002}. Galaxies are fitted with either a S\'ersic + Exponential or S\'ersic + S\'ersic model, with the optimal model selected based on the Bayesian Information Criterion. Classically, LSB galaxies are defined by the disk central surface brightness in the $B$ band, $\mu_{\rm 0,d}(B) \ge 22.5$ mag arcsec$^{-2}$ \citep{McGaugh_s_1996, Rosenbaum_2009, Du_2015}, approximately one magnitude fainter than the Freeman value. Since our bulge–disk decomposition is performed in the $g$ band, we adopt an equivalent threshold of $\mu_{\rm 0,d}(g)=22\pm0.3$ mag arcsec$^{-2}$, accounting for color transformations and typical surface brightness uncertainties. Based on this threshold, we define three subsamples: LSB (159), LSB candidates (388), and HSB (571) galaxies (see details in \citet{Shen_2026}). Because LSB candidates may include both LSB and HSB galaxies, this study focuses on the LSB and HSB samples. The median redshifts for LSB and HSB sample are z = 0.033 and z = 0.025, respectively.
    
    Global stellar masses ($M_\ast$) were taken from the GSWLC-A2 catalog, with the SFR derived from UV/optical SED fitting \citep{Salim_2016, Salim_2018}. Other intrinsic properties, such as redshift ($z$), effective radius ($R_e$), axis ratio ($q = b/a$), and photometric major axis position angle ($\phi$), were obtained from the NASA-Sloan Atlas \citep{Blanton_2011}. Gas-phase metallicity (12+log(O/H)) within one $R_e$ was measured from integrated MaNGA spectra \citep{Westfall_2019}. We further cross-match with the MaNGA DynPop catalog to obtain the look-back times when galaxies formed 50\% and 90\% of their stellar mass (T$_{50}$, T$_{90}$; \citeauthor{Lu_2023} \citeyear{Lu_2023}) and the beam-corrected specific stellar angular momentum within 1 $R_e$ isophote ($\lambda_{\rm Re, \ast}$, \citeauthor{Zhu_Lu_2023} \citeyear{Zhu_Lu_2023}). Stellar velocity (V$_\ast$) and velocity dispersion ($\sigma_\ast$) were taken from the DAPall catalog \citep{Westfall_2019}. The ratio $V_\ast/\sigma_\ast$ is computed using inclination-corrected stellar rotation velocities based on the observed axial ratio ($b/a$). Morphological parameters of \citet{Ye_2025} provided the probabilities of bar (P$_{\rm bar}$), spiral arms (P$_{\rm spiral}$), and merger (P$_{\rm merge}$), with the combined bar+spiral probability defined as P$_{\rm spiral+bar}$ = P$_{\rm spiral}$ + P$_{\rm bar}$ – P$_{\rm spiral}$ × P$_{\rm bar}$. Finally, H$\,\textsc{i}$ masses ($M_{H\,\textsc{i}}$) were obtained by cross-matching with the H$\,\textsc{i}$–MaNGA catalog \citep{Masters_2019}. 
    
    \subsection{Measurements of Multiscale Environments, SFR and 12+log(O/H)}
    \label{sec:Measurements of Multiscale Environments, SFR and 12+log(O/H)}
    
    We characterize galaxy environments using the multiscale framework of \citet{Yesuf2022}. Stellar mass overdensities are computed within fixed physical apertures of radii $x =$ 0.1, 0.25, 0.5, 1, 2, 4, and $8\,h^{-1}\,\mathrm{Mpc}$. For each galaxy, we sum the stellar mass of neighbors within a given aperture and velocity offset $|\Delta v| < 1000$ km s$^{-1}$ and normalize by the median stellar mass density of MaNGA galaxies at the same redshift (in bins of $\Delta z = 0.02$). This yields overdensities $1+\delta_{x,\mathrm{Mpc}/h}$, tracing environments from close surroundings ($0.1\,h^{-1}\,\mathrm{Mpc}$) to the cosmic web ($8\,h^{-1}\,\mathrm{Mpc}$).
    
    To complement these fixed-aperture measures, we compute nearest-neighbor statistics—especially the first-nearest-neighbor distance—as a sensitive probe of close pairs and galaxy isolation. Central and satellite galaxies are identified using the SDSS group catalog \citep{Yang_2007}, which assigns group membership and halo masses iteratively, defining the most massive galaxy as the centrals and the rest as satellites. To verify the robustness of this classification, we conducted a test using the central and satellite galaxy classifications from \citet{Tinker_2020, Tinker_2021}, and found that the results were consistent with those from \citet{Yang_2007}.
    
    Spectroscopic incompleteness in SDSS is mitigated by incorporating DESI DR1 spectroscopic and photometric catalogs \citep{Zou+22,Zou+24}. For photometric neighbors, we test the robustness by varying the velocity window to $|\Delta v| =$ 3000, 6000, and 9000 km s$^{-1}$. Together, these multiscale metrics, fixed aperture overdensities, nearest-neighbor distances, and central/satellite classifications, provide a comprehensive view of environments from small-scale interactions to large-scale structure. Details of the DESI photometric-redshift data will be presented in a companion paper (Yesuf, in prep.).
    
    Applying these methods, we derive environmental parameters for 142 LSB and 516 HSB galaxies, which were cross-matched with the \citet{Yang_2007} group catalog to classify centrals and satellites. Among LSB galaxies, 113 of 142 (80\%) are centrals and 29 are satellites, while for HSB galaxies, 374 of 516 (72\%) are centrals and 142 are satellites. A chi-square test ($\chi^2$(1) = 4.77, p = 0.029) confirms that LSB galaxies host a slightly higher fraction of centrals than HSB galaxies.
    
    We derive SFRs from the dust-corrected H$\alpha$ luminosities ($L_{H\alpha}$) measured by MaNGA. Dust attenuation $A(H\alpha)$, is estimated from the Balmer decrement assuming an intrinsic H$\alpha$/H$\beta$ ratio of 2.86 and the Galactic extinction curve with $R_V$ = 3.1 \citep{Cardelli_1989}. The intrinsic H$\alpha$ luminosity yields the SFR following \citet{Kennicutt_1998} for a Chabrier IMF \citep{Chabrier_2003}:
    \begin{equation}
        \begin{aligned}
            SFR({\rm M_\odot/yr})=4.68 \times 10^{-42} L_{H\alpha}({\rm erg/s})
            \label{formula1}
        \end{aligned}
    \end{equation}
    
    The spatially resolved SFR surface density ($\Sigma_{\rm SFR}$)is obtained by dividing the SFR of each spaxel by its physical area, considering only spaxels with H$\alpha$ signal-to-noise ratios (S/N) $\geq 5$.
   
    Gas-phase metallicities, key tracers of star formation history, are derived from MaNGA emission-line fluxes. We adopt the R23 calibration of \citet{Tremonti_2004}:
    \begin{equation}
        \begin{aligned}
            12+\log {\rm(O/H)}=9.185-0.313x-0.264x^2-0.321x^3
            \label{formula2}
        \end{aligned}
    \end{equation}
    Where $x$ = log(([O$\,\textsc{ii}$]$\lambda\lambda$3726, 3729 + [O$\,\textsc{iii}$]$\lambda\lambda$4959, 5007)/H$\beta$), with [O$\,\textsc{ii}$] and [O$\,\textsc{iii}$] represent the summed fluxes of their respective emission lines. 
    
    For comparison, we also compute metallicities using the $R$ calibration of \citet{Pilyugin_2016} (PG16) and the N2S2H$\alpha$ method of \citet{Dopita_2016} (DOP16), both less sensitive to ionization \citep{Hwang_2019}. Metallicities from PG16 and DOP16 are typically $\sim$0.4 dex lower than those from R23, consistent with differences between theoretical (R23) and direct-method (PG16, DOP16) calibrations. Only spaxels with S/N $\geq 5$ are included in all estimates.

    We construct radial profiles for $\Sigma_{\ast}$, star formation rate surface density ($\Sigma_{\rm SFR}$), specific star formation rate (sSFR), 12+log(O/H), $D_{n}$4000, and H$\delta_A$. To account for galaxy inclination and ensure physical radial distances, we define concentric elliptical rings based on the galaxy's axial ratio ($b/a$) and position angle ($\phi$), with widths of 0.2 $R_e$ from 0.0 to 1.4 $R_e$. The median value of each parameter is calculated within each ring. Due to MaNGA's non-uniform spatial coverage, with most galaxies observed within $\sim$1.5 $R_e$, we limit our analysis to radii $\leq$1.4 $R_e$.
    
    \subsection{Control Sample of LSB and HSB Galaxies}
    \label{sec:Control Sample of LSB and HSB Galaxies}
    
    Based on the LSB galaxy sample from \citet{Shen_2026}, LSB galaxies are predominantly low-mass systems ($M_\ast < 3 \times 10^{10}$ M$_\odot$), with most distributed between 10$^{9}$–10$^{10}$ M$_\odot$. We therefore restrict our analysis within this stellar mass range, yielding 108 LSB galaxies and 211 HSB galaxies.
    
    To enable direct comparison, we construct a stellar mass– and environment-matched subsample using a multivariate nearest-neighbor approach. Each LSB galaxy is paired with an HSB galaxy that satisfies $|\Delta \log M_{\ast}| \leq 0.1$ dex and $|\Delta \log(1+\delta_{1,2,4,8,\mathrm{Mpc}/h})| \leq 0.3$ dex, where log(1+$\delta_{1,2,4,8,\mathrm{Mpc}/h}$) are the multiscale environmental measures described in Section \ref{sec:Measurements of Multiscale Environments, SFR and 12+log(O/H)}. We also considered matching first-nearest-neighbor distance(log(r$_1$)), but the results were similar to those based on the above environmental parameters. Due to the smaller sample size after matching log(r$_1$), we present the results from matching log(1+$\delta_{1,2,4,8,\mathrm{Mpc}/h}$) and log(M$_\ast$). This matching procedure ensures that observed differences reflect intrinsic galaxy properties rather than global scaling relations. After matching, we yield 60 central LSB galaxies matched with 73 central HSB galaxies, while 25 central LSB galaxies remain unmatched. Among satellites, only 7 LSB galaxies match 9 HSB galaxies—insufficient for robust statistics—so satellite galaxies are excluded from environment-based analysis. For completeness, Appendix~\ref{sec:Stellar Mass Matched Results} presents results based on mass-only matching, which provides a larger sample of 85 central LSB galaxies matched with 146 central HSB galaxies, and 23 satellite LSB galaxies matched with 65 satellite HSB galaxies. To ensure a fair statistical comparison of the radial profiles between the matched LSB and HSB galaxy samples, we applied a counting weighting method, which accounts for the sample size imbalance by matching each LSB galaxy with a corresponding HSB galaxy.

    \section{RESULTS}
    \label{sec:RESULTS}

    \subsection{The Multiscale Environments of LSB and HSB galaxies}
    \label{sec:The Multiscale Environments of LSB and HSB galaxies}
    
\begin{figure*}[htbp]%
    \centering
    \includegraphics[width=0.98\textwidth]{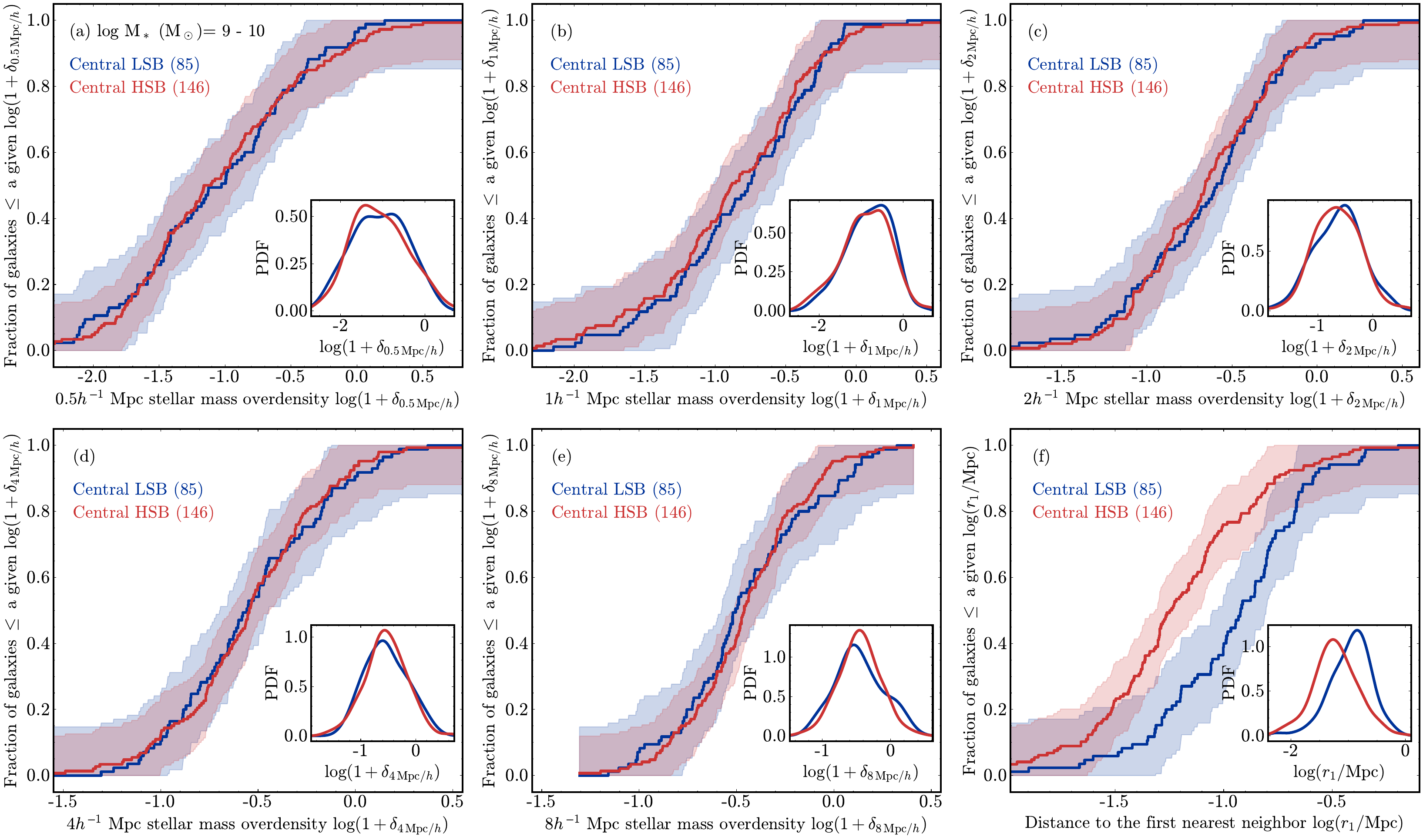}
    \caption{The ECDFs and PDFs (in the inset) of the normalized stellar mass overdensity for central LSB (blue) and HSB (red) galaxies, measured at scales of $\log(1+\delta_{0.5, 1, 2, 4, 8 \,\mathrm{Mpc}/h})$ and for the first nearest neighbor distance (in Mpc). Shaded areas indicate the DKW ($\alpha=0.05$) confidence bands for the ECDFs. Multiscale environments of central LSB and HSB galaxies are remarkably similar, differing only on scales $\lesssim 100~\mathrm{kpc}$. }
    \label{fig1}
    \end{figure*}

    Figure~\ref{fig1} presents the empirical cumulative distribution functions (ECDFs) and probability density functions (PDFs) of the normalized stellar mass overdensity for central LSB and HSB galaxies, measured on scales of $\log(1+\delta_{0.5, 1, 2, 4, 8\,\mathrm{Mpc}/h})$ and for the first nearest neighbor distance (in Mpc). Blue and red lines correspond to LSB and HSB galaxies, respectively, with shaded regions denoting the Dvoretzky–Kiefer–Wolfowitz (DKW,$\alpha=0.05$) confidence bands for the ECDFs. As quantified by Kolmogorov–Smirnov (KS) tests, the multiscale environments of central LSB and HSB galaxies are statistically indistinguishable across all scales up to $8\,\mathrm{Mpc}/h$ (KS statistics $D = [0.08,\,0.08,\,0.08,\,0.08,\,0.09,\,0.07,\,0.11]$ with $p = [0.88,\,0.81,\,0.87,\,0.81,\,0.73,\,0.93,\,0.48]$ for $\delta_{x\,\mathrm{Mpc}/h}$ at $x = 0.1,\,0.25,\,0.5,\,1,\,2,\,4,\,8$, respectively). The environmental number densities based on third- and fifth-nearest neighbors are also consistent between central LSB and HSB galaxies (not shown). In contrast, first-nearest-neighbor distances differ significantly between the two populations ($D = 0.40$, $p < 10^{-8}$), with LSB galaxies being relatively more isolated, with half lacking any neighbor within $\sim 150$ kpc. This result remains robust when neighbors are defined using photometric redshifts, mitigating potential spectroscopic incompleteness in SDSS.
    
\begin{figure*}[htbp]%
    \centering
    \includegraphics[width=0.98\textwidth]{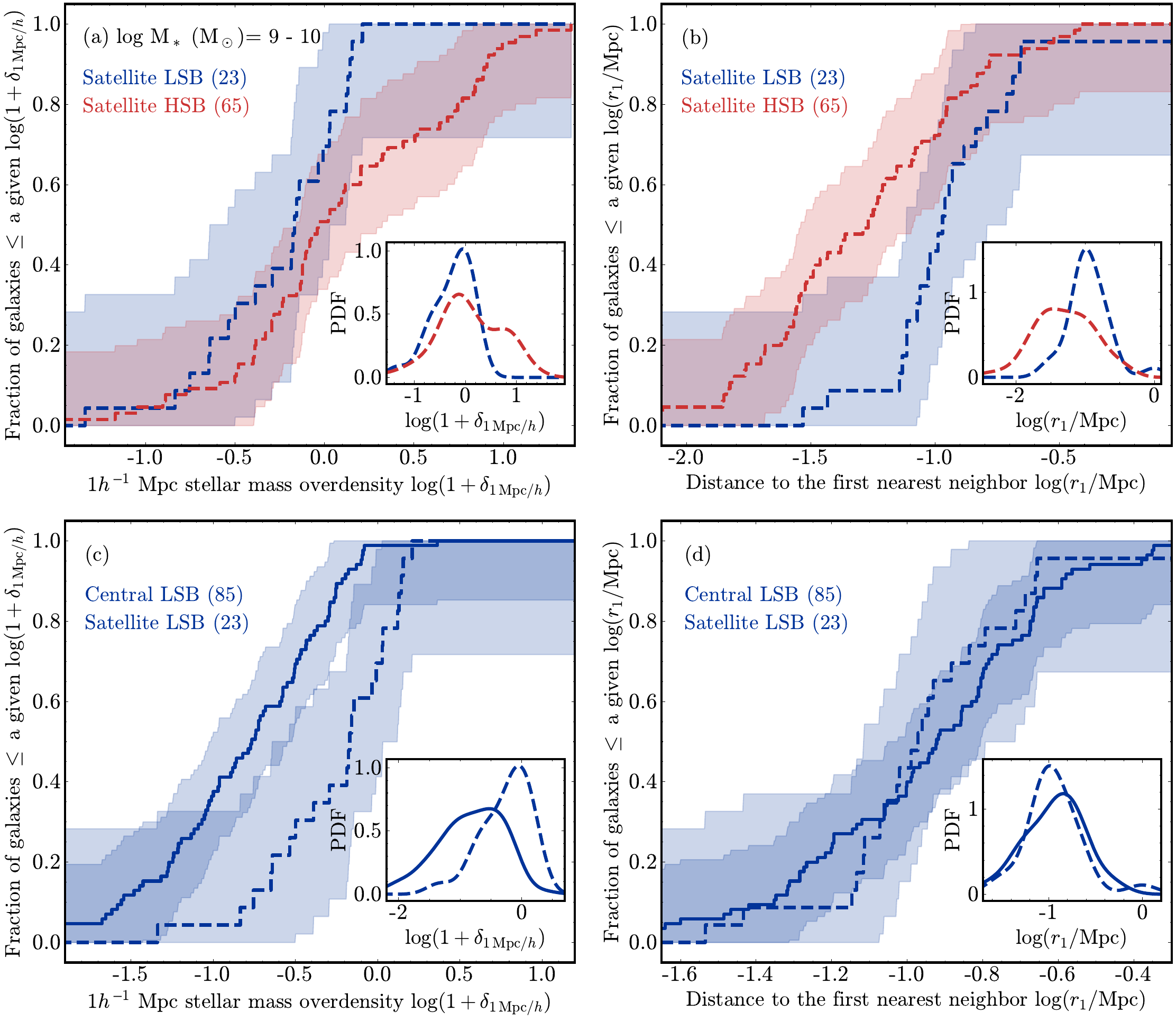}
    \caption{The mass overdensity within 1 Mpc/$h$ and the first nearest-neighbor distances for satellite LSB and HSB galaxies (first row). And the same comparison for central and satellite LSB galaxies (second row). Red and blue lines represent LSB and HSB galaxies, respectively, with solid and dotted lines indicating centrals and satellites. Satellite LSB galaxies exhibit significantly denser environments than satellite HSB and central LSB galaxies, especially on scales of $1-4 \,\mathrm{Mpc}/h$.}
    \label{fig2}
    \end{figure*}

    Although the number of satellite LSB galaxies is limited, we find statistically significant differences between satellite LSB and HSB galaxies in both local and large-scale environments. The distributions of the first-nearest-neighbor distance(log(r$_1$)) and the mass overdensity on scales $1$–$8\,\mathrm{Mpc}/h$ differ markedly, with $D = 0.56$ and $p = 1.9 \times 10^{-5}$ for the log(r$_1$), and $D = [0.36,\,0.31,\,0.38,\,0.35]$ with $p = [0.02,\,0.05,\,0.01,\,0.03]$ for $\delta_{x\,\mathrm{Mpc}/h}$ at $x = 1,\,2,\,4,\,8$, respectively. Distributions on scales outside $1$–$4\,\mathrm{Mpc}/h$ are not significant. Comparing central and satellite LSB galaxies, the log(r$_1$) distributions are statistically similar ($D = 0.18$, $p = 0.51$), but the multiscale overdensities show significant differences across multiple scales, with $D = [0.49,\,0.62,\,0.63,\,0.53,\,0.38,\,0.40,\,0.37]$ and $p = [1.9\times 10^{-4},\,4.7\times 10^{-7},\,2\times 10^{-7},\,3.7\times 10^{-5},\,6.4\times 10^{-3},\,4.3\times 10^{-3},\,0.011]$ for $\delta_{x\,\mathrm{Mpc}/h}$ at $x = 0.1,\,0.25,\,0.5,\,1,\,2,\,4,\,8$, respectively. For illustration, Figure~\ref{fig2} compares the first-nearest-neighbor distance and the mass overdensity within $1\,\mathrm{Mpc}/h$ for all four populations.
    
    \subsection{Stellar Mass and Environment Matched Results}
    \label{sec:Stellar Mass and Environment Matched Results}
    
    This section presents the main results of this study. After controlling for stellar mass and multiscale environments, central LSB galaxies remain systematically distinct from HSB galaxies in their structural, star-forming, and stellar population properties, both globally and in their resolved radial profiles.
    
    \subsubsection{Comparison in Global Properties}
    \label{sec:Comparison in Global Properties}

    This section shows that, after matching both M$_\ast$ and multiscale environments, central LSB galaxies still differ systematically from central HSB galaxies. Specifically, central LSB galaxies exhibit lower SFRs, lower gas-phase metallicities, and larger effective radii. They also show smaller T$_{50}$ and T$_{50}$-T$_{90}$, larger $\lambda_{\rm Re, \ast}$ and $V_\ast/\sigma_\ast$, as well as higher P$_{\rm spiral+bar}$ and lower P$_{\rm merge}$.
    
\begin{figure*}[htbp]%
    \centering
    \includegraphics[width=0.98
    \textwidth]{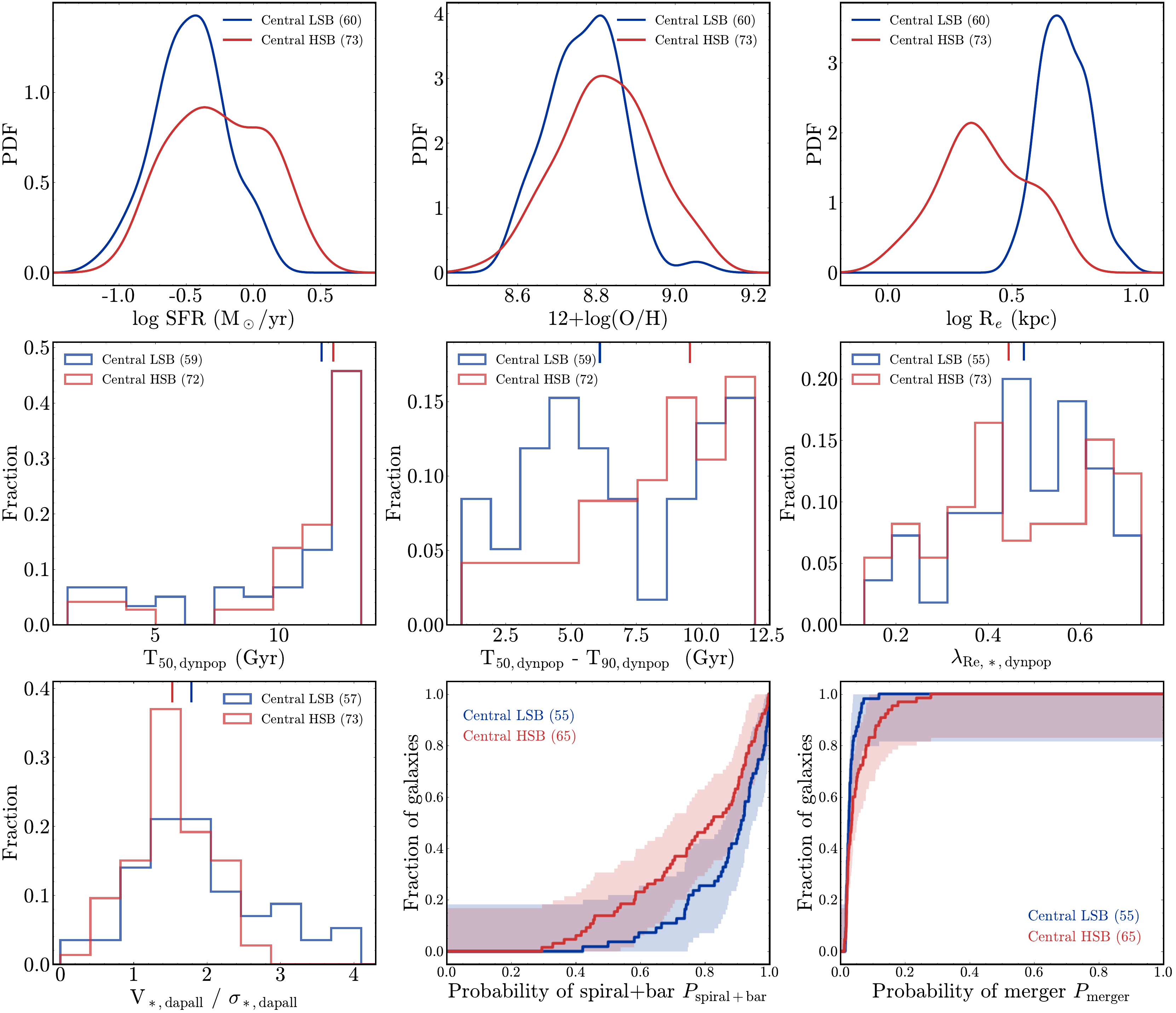}
    \caption{The normalized PDFs of SFR, 12+log(O/H) and size (first row); the histograms of the assembly look-back times for 50\% of the current stellar mass, T$_{50}$, T$_{50}$-T$_{90}$, and specific stellar angular momentum, $\lambda_{\rm Re, \ast}$ (second row); and the histograms of rotation to velocity dispersion ratios, V$_\ast$/$\sigma_\ast$, along with the ECDFs of spiral+bar probabilities, P$_{\rm spiral+bar}$, and merger probabilities, P$_{\rm merge}$ (third row) for central LSB and HSB galaxies after matching M$_\ast$ and multiscale environments. Blue and red solid lines denote central LSB and HSB galaxies, respectively. Median values of each parameter are indicated above the corresponding histograms. Central LSB galaxies are intrinsically distinct from central HSB galaxies, with lower SFRs and metallicities, larger sizes, more extended SFHs, colder kinematics, and higher spiral/bar probabilities but fewer mergers.}
    \label{fig3}
    \end{figure*}

    The first row of Fig.~\ref{fig3} shows that central LSB galaxies have systematically lower SFRs, lower 12+log(O/H), and larger $R_e$ than their HSB counterparts. K–S tests confirm that these differences are highly significant (SFR: $D=0.32$, $p=1.4\times10^{-3}$; 12+log(O/H): $D=0.30$, $p=3.9\times10^{-3}$; $R_e$: $D=0.78$, $p=6.1\times10^{-22}$). These findings indicate that the distinct properties of LSB galaxies—suppressed star formation, reduced chemical enrichment, and extended sizes—are primarily intrinsic in origin rather than environmentally driven.
    
    The second row of Fig.~\ref{fig3} compares their star formation histories using the look-back times T$_{50}$ and T$_{50}$–T$_{90}$. Central LSB galaxies show more extended star formation histories, with a statistically significant difference in T$_{50}$–T$_{90}$ ($D=0.29$, $p=6.5\times10^{-3}$), but not in T$_{50}$ ($D=0.17$, $p=0.25$).

    Further comparison of their stellar kinematic properties reveals that central LSB galaxies have slightly higher $\lambda_{\rm Re,\ast}$ ($D=0.19$, $p=0.19$) and significantly higher $V_\ast/\sigma_\ast$ ($D=0.27$, $p=0.01$), indicating that they are more rotation-dominated and dynamically colder than central HSB galaxies, consistent with a more prominent disk-like structure.

    Finally, the third row of Fig.~\ref{fig3} presents their morphological parameters, P$_{\rm spiral+bar}$ and P$_{\rm merge}$. Central LSB galaxies have significantly higher P$_{\rm spiral+bar}$ ($D=0.27$, $p=0.02$) but lower P$_{\rm merge}$ ($D=0.27$, $p=0.02$), suggesting that they are more likely to host spiral arms or bar structures—reflecting higher internal structural complexity—while being less affected by mergers or interactions. In Appendix \ref{sec:The Image of LSB and HSB Galaxies}, we further present images of central LSB and HSB galaxies matched in M$_\ast$ and multiscale environments to visually illustrate these morphological differences.
    
    \subsubsection{Comparison in Radial Profiles and Gradients}
    \label{sec:Comparison in Radial Profiles and Gradients}
    
\begin{figure*}[htbp]%
    \centering
    \includegraphics[width=0.98\textwidth]{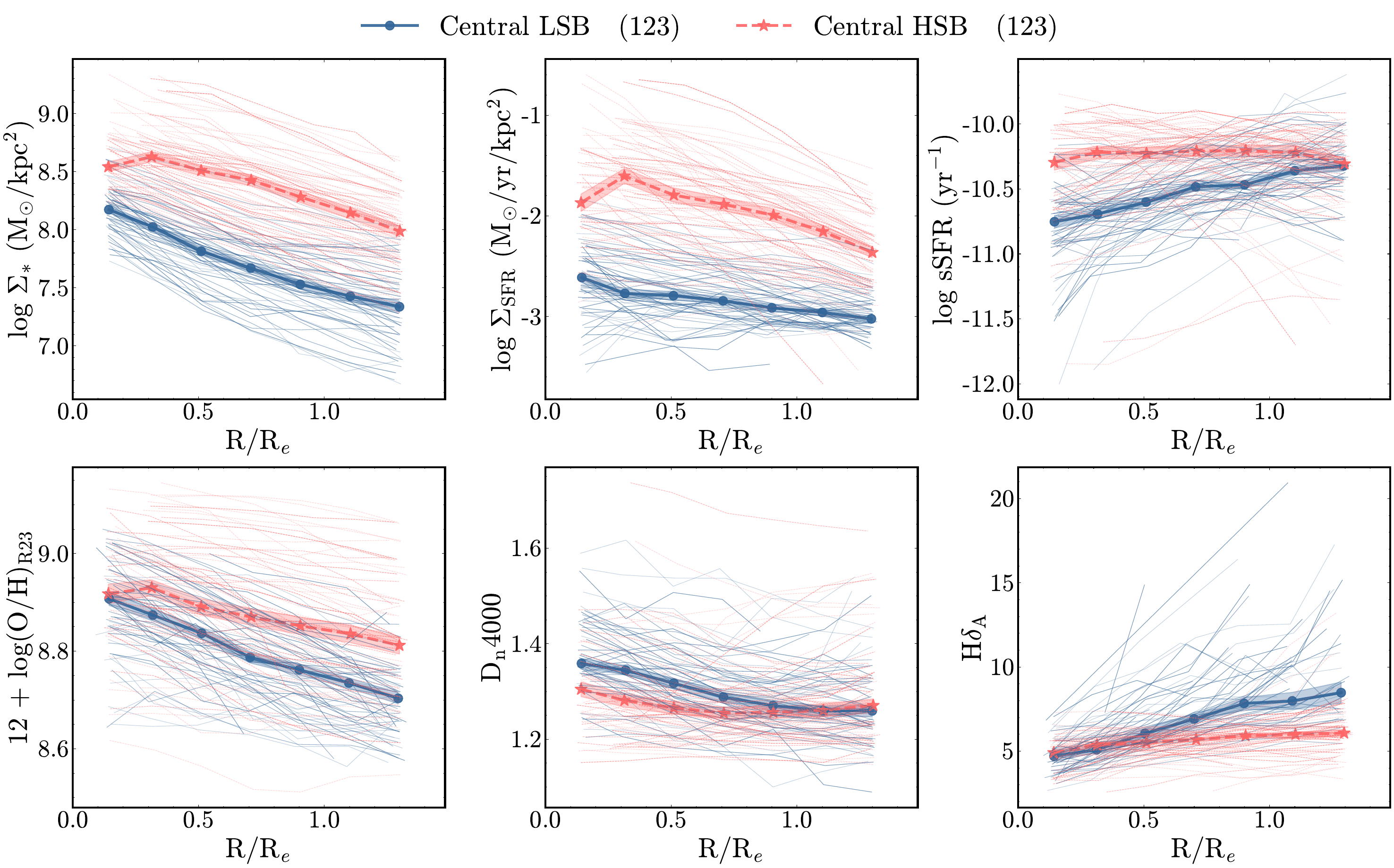}
    \caption{Radial profiles of $\Sigma_\ast$, $\Sigma_{\rm SFR}$, sSFR, 12+log(O/H)$_{\rm R23}$, 4000~\AA{} break strength ($D_n$4000), and Balmer absorption index (H$\delta_A$) as functions of normalized radius ($R/R_e$) for central LSB and HSB galaxies after matching in $M_\ast$ and multiscale environments. Blue solid lines and red dashed lines denote individual central LSB and HSB galaxies, respectively. Blue filled circles and red dotted stars show the median radial profiles of the central LSB and HSB samples. The shaded regions indicate the uncertainty of the median profiles, estimated as $\bar{\sigma} = \sigma / \sqrt{n}$, where $\sigma$ is the standard deviation and $n$ is the sample size. The control matching demonstrates that the visibly distinct radial profiles of central LSB and HSB galaxies are not driven by stellar mass, environmental effects, or, by extension, halo mass.}
    \label{fig4}
    \end{figure*}

    Similarly, after matching by M$_\ast$ and multiscale environments, central LSB galaxies remain systematically distinct from HSB galaxies in both their radial profiles and gradients (Figure~\ref{fig4}). Given the MaNGA PSF, the derived radial profiles should be interpreted as smoothed, resolution-limited trends rather than fully independent spatial measurements. LSB galaxies show lower $\Sigma_{\ast}$, suppressed $\Sigma_{\rm SFR}$, reduced metallicities, and distinct stellar populations—older central stars (higher $D_n$4000) and enhanced recent star formation on their outskirts (stronger H$\delta_A$). Their gradients reinforce these trends (Table \ref{table1}): LSB galaxies exhibit steeper negative $\Sigma_{\ast}$ and metallicity gradients, flatter $\Sigma_{\rm SFR}$ gradients, positive sSFR gradients, and steeper negative $D_n$4000 and H$\delta_A$ gradients. These results suggest that LSB galaxies follow intrinsically different evolutionary pathways from HSB galaxies. The flatter $\Sigma_{\rm SFR}$ and positive sSFR gradients indicate extended, disk-wide star formation and slower chemical enrichment, consistent with inefficient star formation and weak feedback. In contrast, HSB galaxies show more centrally concentrated growth and rapid enrichment. Although environmental processes may modulate these trends, the persistence of observed differences after matching in mass and environment suggests that the dominant drivers are internal, linked to high angular momentum, diffuse baryonic structure, and prolonged gas infall on the outskirts or inefficient radial gas transport within the disk.
    
    \begin{table*}[ht]
    \begin{center}
        \caption{The best-fitting coefficients ($\beta$) in the linear relation \(y = \alpha + \beta x\) for different radial profiles.}
        \renewcommand{\arraystretch}{1.3}
        \setlength{\tabcolsep}{5pt}
        \begin{tabular}{lcccccc}
            \hline\hline 
            \textbf{Galaxy Type} & \textbf{$\Sigma_{\ast}$} & \textbf{$\Sigma_{\rm SFR}$} & \textbf{sSFR} & \textbf{12+log(O/H)} & \textbf{D$_{n}$4000} & \textbf{H$\delta_{A}$} \\
            \hline 
            Central LSB & -0.76$\pm$0.02 & -0.28$\pm$0.02 & 0.34$\pm$0.03 & -0.2$\pm$0.01 & -0.09$\pm$0.01 & 5.64$\pm$0.48 \\
            Central HSB & -0.68$\pm$0.02 & -0.73$\pm$0.04 & -0.02$\pm$0.04 & -0.1$\pm$0.01 & -0.01$\pm$0.01 & 0.81$\pm$0.15 \\
            \hline 
        \end{tabular}%
        \vspace{2mm} 
        \footnotesize 
        \item \textit{Note.} These gradients are measured for the matched samples shown in Fig.~\ref{fig4}.
        \label{table1}
    \end{center}
    \end{table*}
    
    \section{DISCUSSION}
    \label{sec:DISCUSSION}
    
    \subsection{Differences in Environments Between LSB and HSB galaxies}
    \label{sec:Differences in environments between LSB and HSB galaxies}
    
    Our study provides new insights into the long-standing debate on the environments of LSB galaxies. While many earlier studies suggested that LSB galaxies preferentially inhabit low-density environments \citep{Mo_1994, Rosenbaum_2009, Galaz_2011}, our analysis shows a more nuanced picture. For galaxies with 9 $<$ log M$_\ast <$ 10, central LSB galaxies exhibit larger nearest-neighbor distances on small scales ($\sim$100 kpc), indicating greater isolation, but show no significant differences from central HSB galaxies on large scales ($>$100 kpc$\sim$10 Mpc, see Section \ref{sec:The Multiscale Environments of LSB and HSB galaxies}). This result aligns with \citet{Zaritsky_1993} and \citet{Galaz_2011}, but we do not agree with the stronger isolation on 2–5 Mpc scales reported by \citet{Rosenbaum_2009}.
    
    This discrepancy arises mainly from two factors: the scale of environmental measurement and the lack of distinction between central and satellite galaxies in earlier studies. Mixing centrals and satellites introduces bias, since satellite LSB galaxies are more easily disrupted in dense environments and thus underrepresented, giving the false impression that all LSB galaxies prefer isolation. By explicitly separating centrals and satellites, we demonstrate that, at fixed stellar mass, central LSB and HSB galaxies occupy statistically similar large-scale environments, consistent with the broader trend that low-mass galaxies in general reside in average or below-average density regions.
    
    The key differences emerge on the halo scale ($\sim$100 kpc), consistent with \citet{Zaritsky_1993} and \citet{Galaz_2011}. For centrals, LSB and HSB galaxies show no significant differences in large-scale environments, but exhibit differences in spatially resolved properties, reflecting internal evolution. In contrast, satellite LSB galaxies reside in more isolated environments, likely influenced by host halo processes such as ram-pressure stripping and tidal interactions. Thus, previously reported environmental differences in LSB galaxies likely arise from satellite-specific processes measured at small scales, rather than from genuinely distinct large-scale environments.
    
    \subsection{The Formation and Evolution of LSB Galaxies}
    \label{sec:The formation and evolution in LSB galaxies}
    
    The results in Section~\ref{sec:Stellar Mass and Environment Matched Results} reveal a consistent pattern: even after matching in $M_\ast$ and multiscale environment, LSB and HSB galaxies remain systematically different. LSB galaxies have larger effective radii, lower $\Sigma_\ast$, lower gas-phase metallicities, and broader $T_{50}$--$T_{90}$ distributions---differences most pronounced among central galaxies. The environmental analysis in Section~\ref{sec:The Multiscale Environments of LSB and HSB galaxies} further shows that LSB and HSB galaxies inhabit statistically similar large-scale environments, differing only on local ($\sim$100 kpc) scales. Together, these findings imply that their fundamental distinctions arise primarily from intrinsic galaxy or halo properties that correlate only weakly with environment, rather than from environmental influences.
 
    Within the $\Lambda$CDM framework, halo properties such as concentration, spin, and formation time are expected to shape galaxy structure and correlate with large-scale environment through halo assembly bias \citep{Gao_2005, Lee_2017, Behroozi_2022}. If this paradigm holds, galaxies with systematically different sizes or surface brightnesses should occupy distinct environments on average. Our results challenge this expectation: among low-mass centrals, the pronounced LSB–HSB contrast in stellar density profiles and stellar populations persists even after matching in both $M_\ast$ and environment across 0.2$\sim$10 Mpc scales. This persistence indicates that neither halo properties nor environment alone can explain the dichotomy.

    Two aspects underscore this discrepancy. First, the indistinguishable large-scale environments of central LSB and HSB galaxies contradict the idea that their structural differences are environmentally driven. Early-forming, high-concentration halos are predicted to cluster more strongly \citep{Gao_2005}, yet we detect no such distinction. Second, halo spin---the most intuitive driver of surface brightness \citep{Mo_1998}---also fails to explain our observations. While simulations predict moderate environmental variations in spin \citep{Bett_2007, Lee_2017}, LSB and HSB galaxies remain offset in radius and surface brightness even after matching environments. The observed size offset exceeds that expected from the weak size--spin correlations in modern simulations \citep{Desmond_2017, Jiang_2019, Somerville_2025}, consistent with our finding of no significant difference in stellar spin. Spin-based models also fail to reproduce the LSB offset in the baryonic Tully--Fisher relation \citep{McGaugh_2021}. Taken together, these results suggest that variations in halo spin do not fully propagate to the stellar component, likely because feedback or angular momentum redistribution alters the baryonic angular momentum budget.
    
    Recent theoretical studies offer divergent perspectives. Simulations such as EAGLE and TNG suggest that neither halo concentration nor spin alone explains the observed diversity in galaxy radii \citep{Desmond_2017, Somerville_2025}, while \citet{Jiang_2019} argue that halo concentration dominates, predicting $R_{\rm gal} \propto c_{200}^{-0.6}$ with minimal spin dependence. \citet{Behroozi_2022} further show that halo concentration and accretion rate correlate with environment in a scale-dependent manner, implying a potential link between galaxy size and large-scale structure. However, after controlling for $M_\ast$, we find no significant correlation between galaxy radius or surface brightness and environment at any scale (Spearman $\rho < 0.3$). Similarly, in SIMBA simulation, \citet{Cui_2021} found that scatter in the stellar-to-halo mass relation tracks stellar assembly time, which itself follows halo assembly. Yet, based on abundance matching, central LSB and HSB galaxies in our sample likely inhabit halos of similar mass and therefore share comparable average assembly histories, consistent with their similar stellar $T_{50}$. Thus, environmental modulation of halo properties is too weak to leave an observable imprint, or that baryonic processes—such as feedback, angular momentum redistribution, and stochastic assembly histories—decouple galaxy size and surface brightness from halo concentration \citep{Brook_2011, Zjupa_2017}.
    
    Notably, LSB galaxies deviate systematically from the average mass--size relation \citep{Shen_2026}, occupying the large-size end below $M_\star < 3 \times 10^{10}$\, M$_\odot$. This implies that our LSB--HSB comparison directly relates to the large--small galaxy dichotomy in previous studies. Hence, our results provide new observational leverage to test and refine models explaining galaxy size diversity. Future work will quantitatively connect the LSB--HSB distinction with the large--small galaxy dichotomy.
    
    More than 68\% of LSB galaxies in our sample exhibit bars and spiral structures. This contrasts with earlier reports of low bar fractions interpreted as evidence for highly stable, dark matter--dominated disks \citep[e.g.,][]{Mihos_1997}. Those studies targeted smaller, shallower samples from the most diffuse systems. Despite their low stellar surface densities, many LSB galaxies host dynamically cold, rotationally supported disks capable of forming non-axisymmetric features. Simulations show that corotating dark matter halos with moderate spin can facilitate bar and spiral formation by absorbing angular momentum from the disc via resonant interactions \citep{Saha_2013}. The high bar and spiral fractions in isolated LSB galaxies therefore point to internal secular processes rather than tidal interactions as the dominant evolutionary driver. \citet{Narayanan_2024} further demonstrates that global spiral patterns in LSB galaxies can arise as transient instabilities driven by the quadrupolar potential of an oblate dark matter halo, persisting for several Gyr with modest pattern speeds, supporting internally driven disk dynamics.
    
    The observed divergence in central stellar surface density ($\Sigma_\ast$) between LSB and HSB galaxies can be explained primarily by recent \textit{in-situ} star formation modulated by internal gas transport, without invoking external accretion. Assuming roughly constant sSFR during the $T_{50}$--$T_{90}$ interval, HSB galaxies—with longer $T_{50}$--$T_{90}$ ($\sim$9.56~Gyr) and higher central sSFR—can increase their central $\Sigma_\ast$ by $\approx$73\%, whereas LSB galaxies, with shorter $T_{50}$--$T_{90}$ ($\sim$6.1~Gyr) and lower central sSFR ($\sim0.5\times\mathrm{sSFR}_{\mathrm{HSB}}$), contribute only $\approx 23$\% to central mass growth. This difference reproduces the observed $\Sigma_{\mathrm{LSB}} / \Sigma_{\mathrm{HSB}} \approx 0.5$. For HSB galaxies with moderate bulge fractions ($B/T \approx 0.13$–$0.52$), the global H\,\textsc{i} reservoir ($f_{\mathrm{H\,\textsc{i}}} \sim 1$) suffices to sustain this growth without significantly depleting the total gas fraction. Based on their lower central metallicities and lower dust absorption \citep{YesufHo2019}, LSB galaxies likely contain less molecular gas in their centers than HSB galaxies (see Fig. \ref{figA3} in Appendix \ref{The mhi and Av LSB and HSB galaxies}).
    
    In LSB galaxies, inefficient inward transport of high-angular-momentum gas suppresses central mass growth while sustaining star formation in the outer disk. This explains their structural characteristics: high bar and spiral fractions (including some lopsided systems) and large $V_{\mathrm{rot, \ast}}/\sigma_\ast$ ratios reflect dynamically cold disks where star formation and instabilities occur predominantly at larger radii. Differences in \textit{in-situ} star formation, angular-momentum–driven gas transport, and disk dynamical stability---potentially modulated by halo spin and shape---jointly shape the radial $\Sigma_\ast$ profiles of LSB and HSB galaxies, producing their systematic contrast at fixed stellar mass and environment.

    Although secondary, environment can influence galaxy evolution under specific conditions. Simulations \citep[e.g.,][]{Somerville_2025} indicate that star formation efficiency is sensitive to environment, particularly for satellites. Consistently, satellite LSB galaxies show stronger environmental effects on SFR and metallicity, indicating that environment can modulate—but not determine—trends on small scales.
    
    In summary, our results provide new empirical constraints on galaxy assembly bias and the coupling between halo and galaxy properties in the low-mass regime. Galaxy size, structure, and star formation history cannot be predicted solely from halo concentration, spin, or environment. Future observations combining weak lensing and satellite kinematics will be critical to directly probe halo properties and test this integrated framework linking halo physics, baryonic processes, and galaxy structure.

    \section{SUMMARY AND CONCLUSION}
    \label{sec:SUMMARY AND CONCLUSION}

    We used a sample from \citet{Shen_2026}, who defined LSB and HSB galaxies in the MaNGA survey via bulge-disk decomposition with a central disk surface brightness criterion of $\mu_{0,d}(g) = 22 \pm 0.3$ mag arcsec$^{-2}$. From this parent sample, we selected late-type galaxies with $9 < \log M_\ast < 10$ and classified them into centrals and satellites. The final sample includes 113 central and 29 satellite LSB galaxies, and 374 central and 142 satellite HSB galaxies, enabling systematic comparison of their global and spatially resolved properties across multiple environments. The main results are:
    
    \begin{itemize}
    \item Environmental analysis from $\sim$100 kpc to 10 Mpc reveals that central LSB and HSB galaxies reside in statistically similar large-scale ($>$100 kpc) environments, typically of average or below-average density. However, central LSB galaxies have significantly larger $k=1$ nearest-neighbor distances. Environment-driven differences between LSB and HSB galaxies emerge clearly on smaller scales ($\sim$100 kpc), particularly for centrals, which inhabit more isolated local environments.
    
    \item In global properties, even after matching both M$_\ast$ and multiscale environments, central LSB galaxies remain systematically different from central HSB galaxies. They show lower SFRs and metallicities, larger effective radii, and diverse recent star formation histories (larger T$_{50}$–T$_{90}$), indicating a mix of rapidly and slowly evolving systems. Kinematically, LSB galaxies are more rotation-dominated and dynamically colder, with higher $V_\ast/\sigma_\ast$ and slightly higher $\lambda_{\rm Re,\ast}$. Morphologically, their higher P$_{\rm spiral+bar}$ and lower P$_{\rm merge}$ indicate predominantly secular internal evolution with minimal merger or interaction influence.
    
    \item Spatially resolved spectroscopy further reveals that, even with matched M$_\ast$ and multiscale environment, LSB galaxies generally exhibit lower $\Sigma_\ast$, $\Sigma_{\rm SFR}$, sSFR, and gas-phase metallicity compared to HSB counterparts. They also show higher central D$_n$4000 and enhanced H$\delta_A$ in their outskirts. The results imply that LSB galaxies evolve mainly through intrinsic processes, perhaps reflecting the high spin of their host halos, which is weakly correlated with current large-scale environment. While environmental effects are secondary, they remain non-negligible, particularly in LSB satellite galaxies, where environmental influences on star formation and chemical evolution are observed.
    
    \item The differences between LSB and HSB galaxies are primarily driven by intrinsic galaxy properties rather than large-scale environmental factors. Although both types occupy similar large-scale environments, LSB galaxies tend to inhabit more isolated local environments, which may contribute to their distinct star formation histories and lower stellar surface densities. While halo spin may influence the LSB/HSB distinction, intrinsic processes such as gas transport, angular momentum redistribution, and secular evolution are more critical in shaping their differences in size, surface brightness, and metallicity.
    
    \end{itemize}
    
    In summary, our study reinterprets LSB galaxies not as rare anomalies formed in isolation, but as a distinct evolutionary pathway for low-mass galaxies, shaped primarily by intrinsic formation mechanisms and secondarily by environmental influences. Future observations with next-generation telescopes that directly probe gas accretion and dark matter halo properties will be crucial for elucidating this layered evolutionary framework.

    \newpage
    
    \section*{Acknowledgement}
    
    We thank the anonymous referee for comments that helped to improve this paper. This work is supported by the National Natural Science Foundation of China (NSFC) under Nos. 12373007, 12422302. This work is supported by National Key R$\&$D Program of China No.2022YFF0503402, and the National Natural Science Foundation of China (NSFC, Nos. 12233005). J.W. acknowledges NSFC grants 12033004, 12333002, 12221003 and the China Manned Space Program with grant no. CMS-CSST-2025-A10. SS thanks research grants from the China Manned Space Project with NO. CMS-CSST-2025-A07, National Natural Science Foundation of China (No. 12141302) and Shanghai Academic/Technology Research Leader (22XD1404200). JY thanks the support of the China Manned Space Program (Grant No. CMS-CSST-2025-A19).
    
    This work makes use of data from SDSS-IV. Funding for SDSS-IV has been provided by the Alfred PSloan Foundation and Participating Institutions. Additional funding towards SDSS-IV has been provided by the US Department of Energy Office of Science. SDSS-IV acknowledges support and resources from the Centre for High-Performance Computing at the University of Utah. The SDSS web site is www.sdss.org.
    
    SDSS-IV is managed by the Astrophysical Research Consortium for the Participating Institutions of the SDSS Collaboration including the Brazilian Participation Group, the Carnegie Institution for Science, Carnegie Mellon University, the Chilean Participation Group, the French Participation Group, Harvard–Smithsonian Center for Astrophysics, Instituto de Astrofsica de Canarias, The Johns Hopkins University, Kavli Institute for the Physics and Mathematics of the Universe (IPMU)/University of Tokyo, Lawrence Berkeley National Laboratory, Leibniz Institut fur Astrophysik Potsdam (AIP), Max-Planck-Institut fur Astronomie (MPIA Heidelberg), Max-Planck-Institut fur Astrophysik (MPA Garching), Max-Planck-Institut fur Extraterrestrische Physik (MPE), National Astronomical Observatory of China, New Mexico State University, New York University, University of Notre Dame, Observatario Nacional/MCTI, The Ohio State University, Pennsylvania State University, Shanghai Astronomical Observatory, United Kingdom Participation Group, Universidad Nacional Autonoma de Mexico, University of Arizona, University of Colorado Boulder, University of Oxford, University of Portsmouth, University of Utah, University of Virginia, University of Washington, University of Wisconsin, Vanderbilt University and Yale University.
    
    \bibliography{ref}{}
    \bibliographystyle{aasjournal}

    \appendix
    \renewcommand\thefigure{\Alph{section}\arabic{figure}}
    \setcounter{figure}{0} 
    
    \section{The Images of LSB and HSB Galaxies}
    \label{sec:The Image of LSB and HSB Galaxies}

\begin{figure*}[htbp]%
    \centering
    \includegraphics[width=0.98\textwidth]{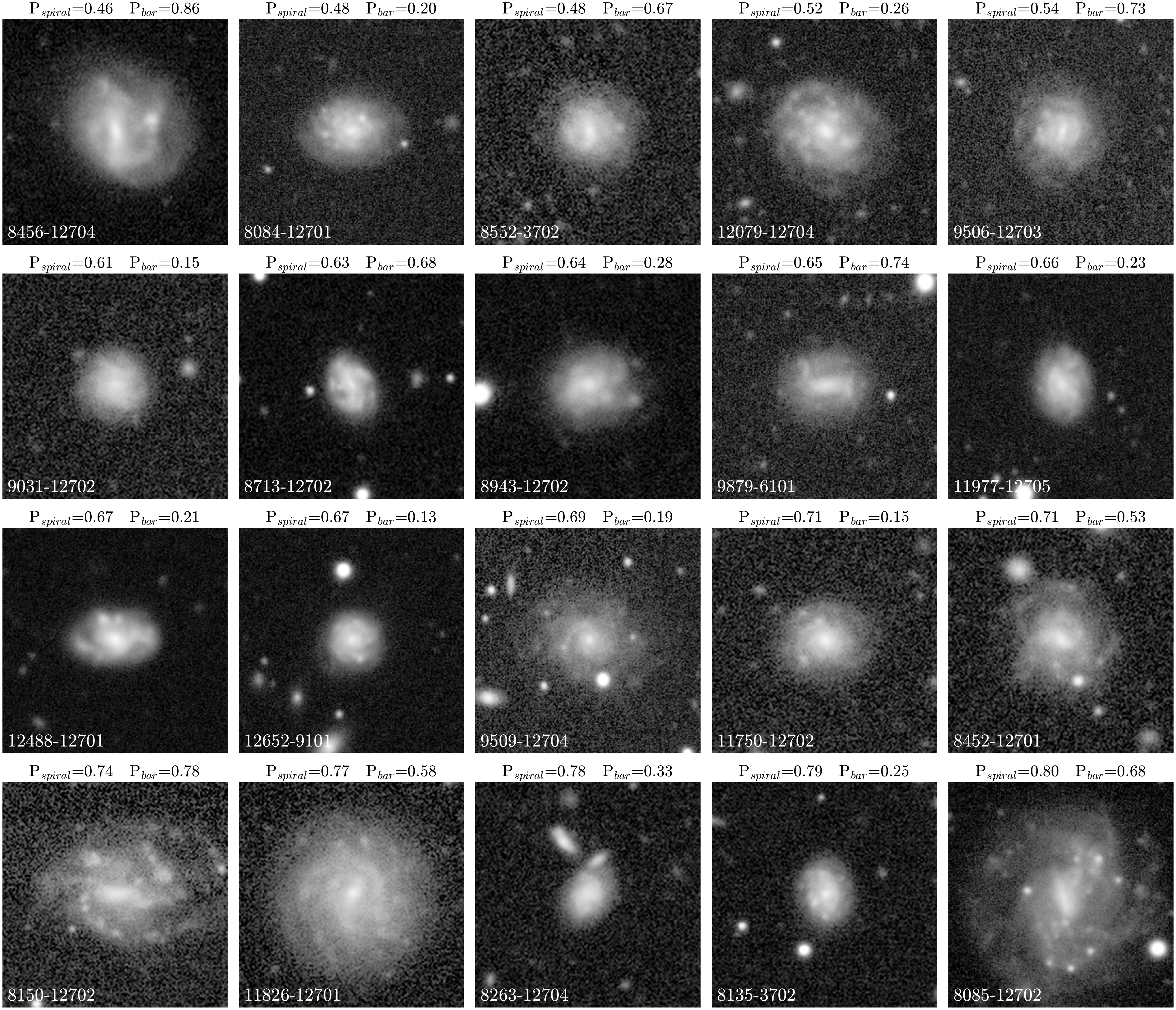}
    \caption{The $g$-band images of 20 central LSB galaxies, ordered by increasing P$_{\rm spiral}$ from left to right. The top of each image indicates the corresponding P$_{\rm spiral}$ and P$_{\rm bar}$ values. These LSB galaxies are matched in mass and environments to HSB galaxies. Each galaxy image is cropped to a size of 240×240 pixel$^2$ (equivalent to 62.88 × 62.88 arcsec$^2$).}
    \label{figA1}
    \end{figure*}

\begin{figure*}[htbp]%
    \centering
    \includegraphics[width=0.98\textwidth]{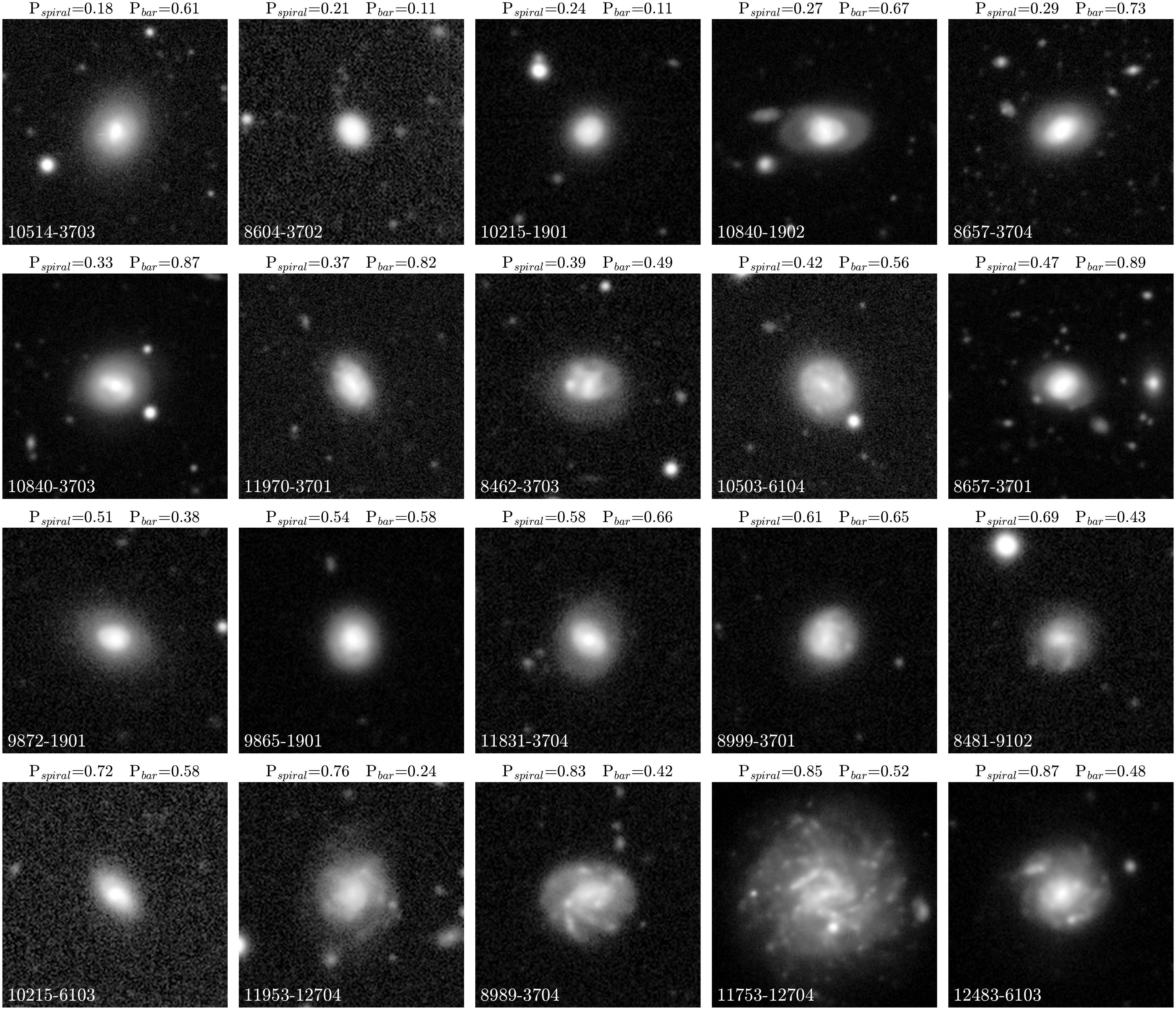}
    \caption{The $g$-band images of 20 central HSB galaxies, ordered by increasing P$_{\rm spiral}$ from left to right. The top of each image indicates the corresponding P$_{\rm spiral}$ and P$_{\rm bar}$ values. These HSB galaxies are matched in mass and environments to LSB galaxies. Each galaxy image is cropped to a size of 240×240 pixel$^2$ (equivalent to 62.88 × 62.88 arcsec$^2$).}
    \label{figA2}
    \end{figure*}

    Figures~\ref{figA1}–\ref{figA2} show $g$-band images of 20 central LSB and HSB galaxies matched in M$_\ast$ and environments. Galaxies are arranged from left to right in order of increasing P$_{\rm spiral}$, with P$_{\rm spiral}$ and P$_{\rm bar}$ values labeled above each panel. LSB galaxies are larger, clumpier, and more likely to show spiral features, reflecting their low surface densities and cold disks, while HSB galaxies are smaller and smoother.

    \section{The distribution of M$_{H\,\textsc{i}}$ and A$_V$ in Central LSB and HSB galaxies}
    \label{The mhi and Av LSB and HSB galaxies}

    We present the distributions of bar probability (P$_{\rm bar}$), H\,\textsc{i}-to-stellar mass ratio (M$_{H\,\textsc{i}}$/M$\ast$), and dust extinction (A$_V$) for mass- and environment-matched central LSB and HSB galaxies. Central LSB and HSB galaxies show similar P$_{\rm bar}$, while LSB galaxies have significantly higher M$_{H\,\textsc{i}}$/M$_\ast$ ($D=0.39$, $p=0.001$) and lower A$_V$ profile.
    Using the formula from \citet{Barrera_2020}, A$_V$ = A$_{\rm H\alpha}$/0.817, the median A$_{\rm H\alpha}$ for central LSB and HSB galaxies is approximately 0.23 and 0.44, respectively.
    
    \begin{figure*}[htbp]%
    \centering
    \includegraphics[width=0.98\textwidth]{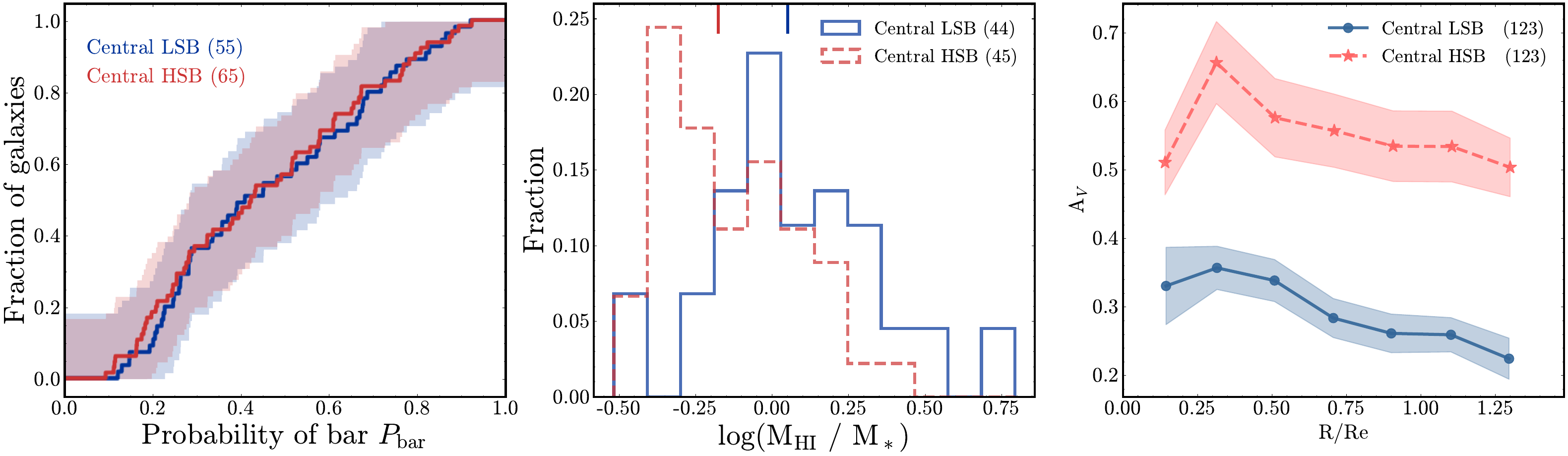}
    \caption{The ECDFs of P$_{\rm bar}$ and the histograms of M$_{H\,\textsc{i}}$/M$_\ast$ and radial profile of A$_V$ for central LSB and HSB galaxies after matching M$_\ast$ and multiscale environments. The color identification is the same as in Fig.~\ref{fig3}.}
    \label{figA3}
    \end{figure*}%
    
    \section{Stellar Mass Matched Results}
    \label{sec:Stellar Mass Matched Results}
    
    This section presents the global properties and radial profiles of mass-matched central and satellite galaxies.
    
    \subsection{Global Properties}
    \label{sec:Global Properties_mass}
    After matching in stellar mass, LSB galaxies—both centrals and satellites—consistently exhibit lower SFRs, lower 12+log(O/H), and larger effective radii than HSB galaxies. The SFR difference is more pronounced among centrals (KS $D=0.35$, $p=2.8\times10^{-6}$) and weaker among satellites ($D=0.28$, $p=0.1$). LSB galaxies are also more metal-poor, with stronger contrasts among satellites ($D=0.55$, $p=2.3\times10^{-5}$) than centrals ($D=0.34$, $p=4.8\times10^{-6}$). Satellite galaxies show slightly higher metallicities than centrals, likely due to environmental processes such as gas stripping and suppressed accretion of metal-poor gas. In size, LSB galaxies are systematically larger than HSB galaxies ($D=0.73$, $p=1.8\times10^{-29}$ for centrals; $D=0.65$, $p=2.7\times10^{-7}$), with LSB centrals being the largest systems overall.

    \begin{figure*}[htbp]%
    \centering
    \includegraphics[width=0.98\textwidth]{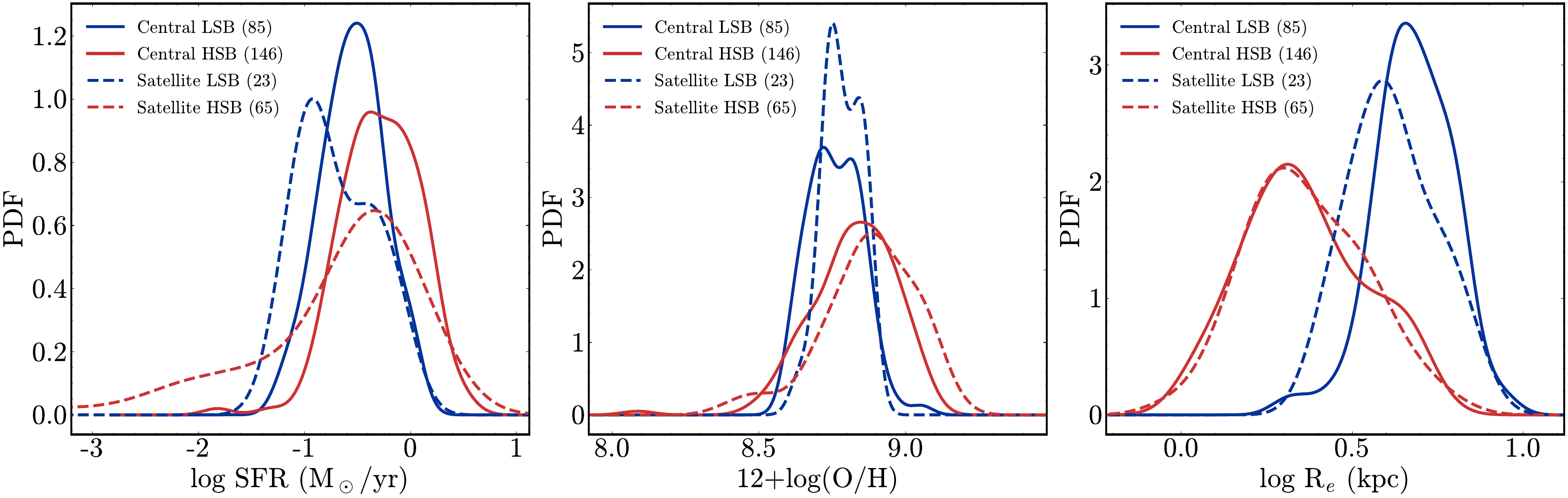}
    \caption{The normalized PDFs of SFR, 12+log(O/H) and R$_e$ for mass-matched central and satellite LSB (blue) and HSB (red) galaxies; solid and dotted lines indicate centrals and satellites, respectively.}
    \label{figA4}
    \end{figure*}%
    
    \subsection{Radial Profiles and Gradients}
    \label{sec:Radial Profiles and Gradients}

\begin{figure*}[htbp]%
    \centering
    \includegraphics[width=0.98\textwidth]{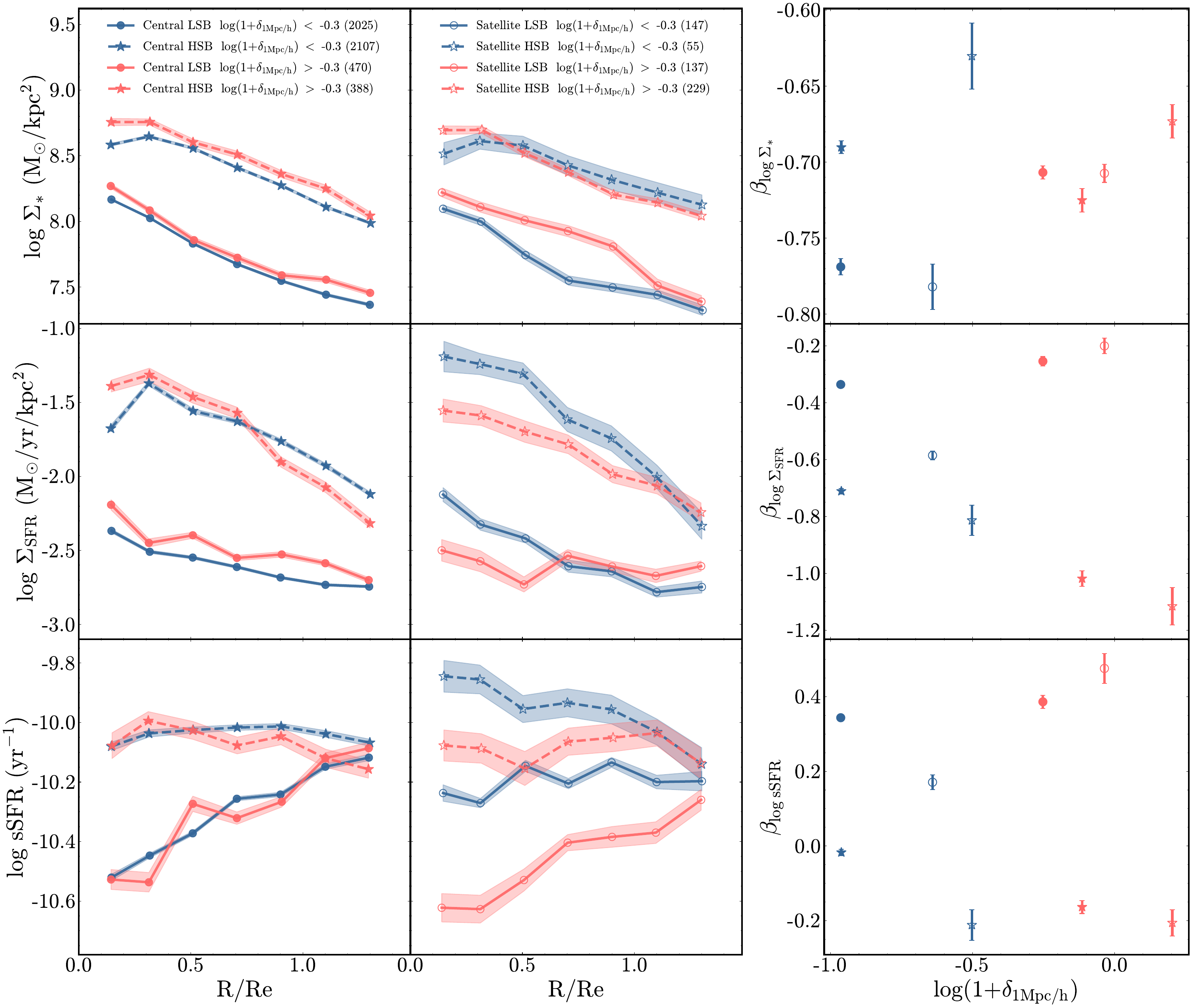}
    \caption{Radial profiles and gradients of $\Sigma_{\ast}$, $\Sigma_{\rm SFR}$, and sSFR for LSB and HSB galaxies after matching M$_\ast$. Rows (top to bottom) show $\Sigma_{\ast}$, $\Sigma_{\rm SFR}$, and sSFR; columns (left to right) display radial profiles for central and satellite galaxies, and their corresponding gradients. Solid circles and dotted stars represent LSB and HSB galaxies, respectively; filled and open symbols indicate centrals and satellites; blue and red denote low- ($\log(1+\delta_{\rm 1Mpc/h}) < -0.3$) and high-density ($\log(1+\delta_{\rm 1Mpc/h}) > -0.3$) environments. Radial profiles show median values with shaded regions indicating errors. Gradient panels use the same symbols and colors, with error bars indicating uncertainties. The legend in the first row lists the number of LSB and HSB galaxies in each environment.}
    \label{figA5}
    \end{figure*}

\begin{figure*}[htbp]%
    \centering
    \includegraphics[width=0.98\textwidth]{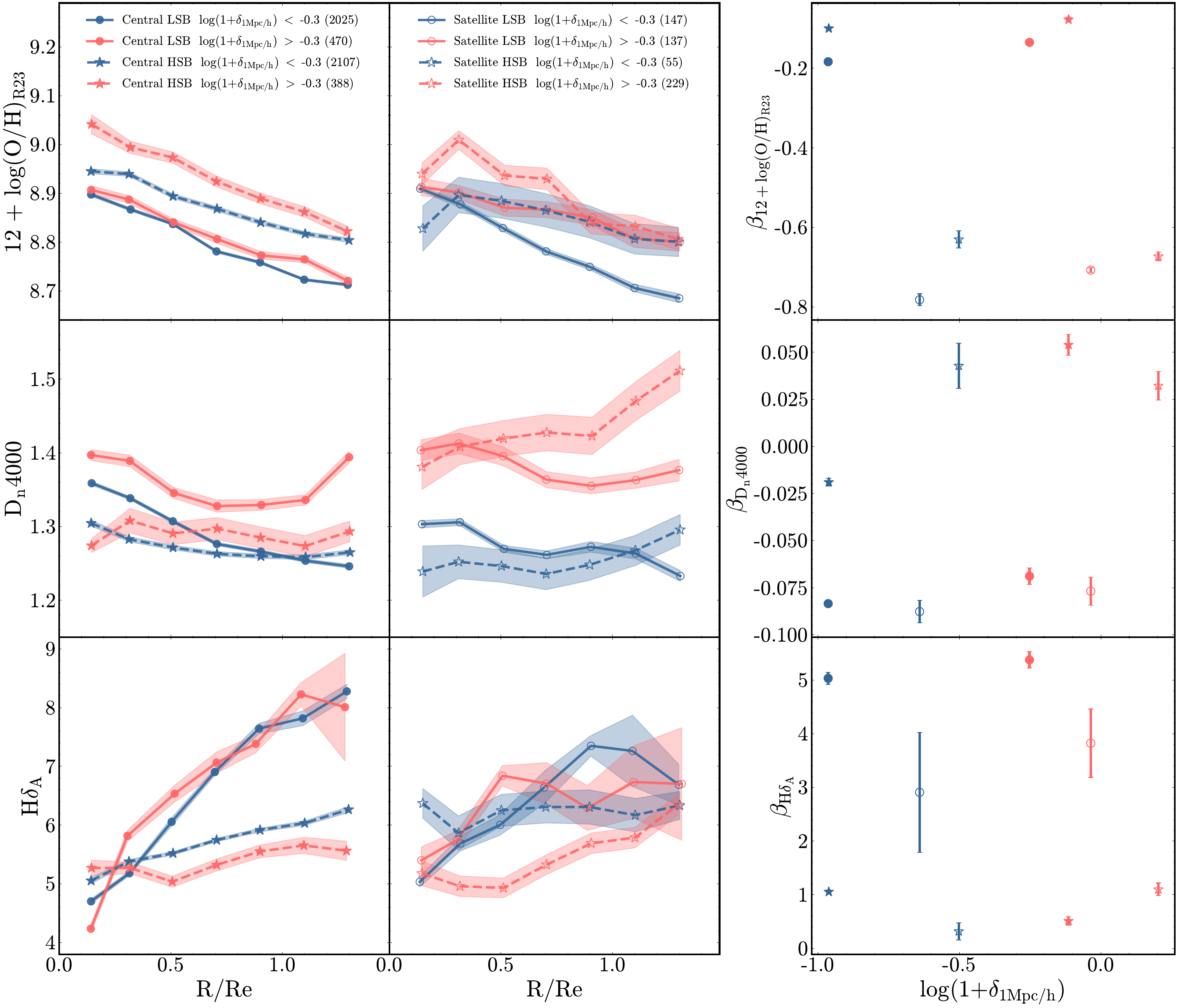}
    \caption{Radial profiles and gradients of 12+log(O/H)$_{\rm R23}$, D$_n$4000 and H$\delta_A$ for different environments after matching M$_\ast$. The color and symbol shapes identification of different environments is the same as in Fig.~\ref{figA5}.}
    \label{figA6}
    \end{figure*}

    Figures~\ref{figA5}–\ref{figA6} present the radial profiles and gradients of $\Sigma_{\ast}$, $\Sigma_{\rm SFR}$, sSFR, 12+log(O/H), $D_{n}4000$, and H$\delta_A$ for mass-matched central and satellite LSB and HSB galaxies in high- and low-density environments (log(1+$\delta_{\rm 1Mpc/h}$)$>-0.3$ and $<-0.3$). 
    
    Figure~\ref{figA5} shows that LSB galaxies have systematically lower $\Sigma_\ast$, $\Sigma_{\rm SFR}$, and sSFR than HSB galaxies, for both centrals and satellites. Their stellar mass and star formation are more extended, with steeper $\Sigma_\ast$ gradients, flatter $\Sigma_{\rm SFR}$ gradients, and positive sSFR gradients, indicating weaker central concentrations and enhanced outer activity. Centrals generally have lower surface densities and flatter slopes than satellites, while satellites are more sensitive to environmental effects. In low-density regions, LSB galaxies show steeper $\Sigma_\ast$ and $\Sigma_{\rm SFR}$ gradients and higher outer sSFR, suggesting that diffuse environments favor extended star formation and slower quenching. High-density regions tend to suppress these differences, likely through processes such as gas stripping or star formation suppression. 

    LSB galaxies are also generally more metal-poor than HSB galaxies, with steeper negative metallicity gradients, particularly for central galaxies in low-density environments, while HSB gradients are relatively insensitive to environment. Central galaxies typically have lower metallicities and flatter gradients than satellites, while environmental effects are stronger for satellites, reflecting processes such as gas stripping or quenching. Stellar population indicators ($D_{n}4000$ and H$\delta_A$) reveal that LSB galaxies have older central populations but younger star-forming outskirts, consistent with extended star formation histories, whereas HSB galaxies are more centrally concentrated and environmentally sensitive. Satellites show more complex and sometimes reversed trends, highlighting the additional impact of environment on their stellar populations. 

\end{CJK}
\end{document}